\documentclass[11pt]{article}
\usepackage[utf8]{inputenc}
\usepackage{amsmath,amssymb}
\usepackage{graphicx}
\usepackage{hyperref}
\usepackage{authblk}
\usepackage{geometry}
\title{Impact of homophily in adherence to anti-epidemic measures on the spread of infectious diseases in social networks}

\author[1]{Piotr Bentkowski}
\author[2]{Tomasz Gubiec}

\affil[1]{University of Warsaw, Artes Liberales; \texttt{p.bentkowski@al.uw.edu.pl}}
\affil[2]{University of Warsaw, Faculty of Physics; \texttt{t.gubiec@uw.edu.pl}}

\date{}

\begin{document}

\maketitle

\begin{abstract}
We investigate how homophily in adherence to anti-epidemic measures affects the final size of epidemics in social networks. 
Using a modified SIR model, we divide agents into two behavioral groups—compliant and non-compliant—and introduce transmission probabilities that depend asymmetrically on the behavior of both the infected and susceptible individuals. 
We simulate epidemic dynamics on two types of synthetic networks with tunable inter-group connection probability: stochastic block models (SBM)  and networks with triadic closure (TC) that better capture local clustering. 
Our main result reveals a counterintuitive effect: under conditions where compliant infected agents significantly reduce transmission, increasing the separation between groups may lead to a higher fraction of infections in the compliant population. 
This paradoxical outcome emerges only in networks with clustering (TC), not in SBM, suggesting that local network structure plays a crucial role. 
These findings highlight that increasing group separation does not always confer protection, especially when behavioral traits amplify within-group transmission.
\end{abstract}

\section{Introduction}
\subsection{Motivation}

The COVID-19 pandemic has sparked renewed interest in understanding how behavioural and structural heterogeneity affect the spread of infectious diseases in social systems. Although classical compartmental models in epidemiology treat the population as homogeneous, real-world societies are composed of individuals with diverse behavioural responses, including varying levels of adherence to public health interventions such as mask wearing, vaccination, and social distancing.

Recent empirical studies have shown that people tend to cluster in social and spatial proximity based on their health behaviours and beliefs \cite{may2003clustering, omer2008geographic, barclay2014positive}, a phenomenon known as homophily. Such clustering can have profound implications for epidemic dynamics, particularly when individuals who neglect anti-epidemic measures form tightly connected subgroups. These clusters can serve as reservoirs for sustained transmission, reducing the overall effectiveness of interventions aimed at the population level.

Our study focusses on understanding how homophily in adherence to antiepidemic measures affects the final size of an epidemic in a population composed of two behavioural groups. We distinguish between compliant individuals who follow preventive measures, and noncompliant individuals who do not. By embedding these groups into synthetic networks with tunable levels of intergroup connectivity, we explore whether increasing separation between behavioural groups protects compliant individuals or, paradoxically, makes them more vulnerable.

We show that under certain conditions, separating groups may inadvertently amplify transmission within the non-compliant cluster, increasing secondary exposure for the compliant group. This finding underscores the importance of considering network structure and asymmetric behavioural effects in the design and evaluation of epidemic control strategies.

\subsection{Related Work}

The role of behavioural clustering in epidemic dynamics has been studied in multiple disciplines, including epidemiology, network science, and computational social science. Early work by May and Silverman \cite{may2003clustering} highlighted the potential threat posed by spatial clusters of individuals who refuse vaccination, showing that such clustering can lower herd immunity thresholds. Omer et al. \cite{omer2008geographic} extended this observation by identifying geographic clustering of non-medical exemptions and its association with pertussis outbreaks.

From a network-theoretic perspective, Salath\'e and Bonhoeffer \cite{salathe2008effect} showed that opinion clustering can dramatically affect outbreak size, even when the overall fraction of protected individuals remains constant. More recent studies have investigated the dual role of social influence and selection in producing clusters of vaccine hesitancy \cite{alvarez2022spatial}, and how assortative mixing based on behaviors such as mask-wearing \cite{watanabe2022impact} or vaccination \cite{hiraoka2022herd, burgio2022homophily, are2024vaccinehomophily} alters epidemic thresholds and dynamics. In particular, \cite{are2024vaccinehomophily} show that vaccine status homophily can amplify infection risk among the unvaccinated, even when overall coverage is high.

The interaction between homophily and the effectiveness of non-pharmaceutical interventions (NPIs) has recently been linked to paradoxical phenomena. Klimek et al. \cite{klimek2024maskshomophily} demonstrate that, under specific behavioral correlations, low mask coverage can be more protective for mask wearers than medium coverage—a result attributed to network-level feedbacks and assortativity. These findings resonate with our observation that increased separation between compliant and non-compliant individuals can elevate risk for the compliant group, depending on which aspect of transmission is modulated by interventions.

From a modelling perspective, our use of networks with controlled clustering and modularity builds on recent advances in network generation. In particular, the STC model proposed in \cite{cirigliano2024stc} enables explicit control of triadic closure while preserving degree distributions. The importance of higher-order motifs has been emphasized in studies of complex contagion \cite{burgio2024triadicapprox, malizia2025gbcm}, showing that clustering and triangle overlap influence both outbreak thresholds and final epidemic size. Bizzarri et al. \cite{bizzarri2024homophilyeffects} theoretically predict that homophily may first amplify and then suppress infections as it increases, depending on epidemic regime.

These studies collectively suggest that homophily in protective behavior is not merely a passive reflection of social preferences but an active driver of epidemic risk. Our work contributes to this literature by demonstrating that, under realistic constraints on transmission asymmetry, increased homophily can produce counterintuitive outcomes. Unlike most prior models, we examine this effect in networks with controlled modularity and clustering, and we isolate the conditions under which increased separation between compliant and non-compliant individuals harms rather than protects the compliant population.

\subsection{Research Questions and Contribution}

This work addresses the following research question: under what conditions does increasing behavioral homophily—interpreted as the separation between compliant and non-compliant individuals—lead to reduced epidemic impact in the compliant group? More specifically, we aim to understand how network structure and asymmetric transmission dynamics influence the final size of the epidemic within each behavioral group.

To answer this question, we develop a modified SIR model in which the probability of transmission depends asymmetrically on the behaviors of both the infected and susceptible individuals. We embed the population into synthetic networks generated by two models which allows control over inter-group connection probabilities: the stochastic block model (SBM) and a triadic closure (TC) model, which captures clustering typical of real-world social networks.

The key contributions of this study are:
\begin{itemize}
    \item We introduce a variant of the SIR model with behavior-dependent transmission asymmetries, distinguishing between the infectiousness of non-compliant individuals and the susceptibility of compliant individuals.
    \item We systematically explore epidemic dynamics across a range of homophily levels and model parameters on both SBM and TC networks, ensuring constant network density for fair comparison (mean degree \( \langle k \rangle = 4 \) )
    \item We identify a counterintuitive effect: in clustered networks (TC), increasing the separation between groups can increase the fraction of recovered nodes in the compliant group, due to intensified intra-group transmission among non-compliant individuals.
    \item We show that this effect does not appear in SBM networks, underscoring the role of local clustering in mediating epidemic outcomes.
\end{itemize}

By isolating the conditions under which homophily exacerbates risk rather than mitigates it, our findings contribute to the theoretical understanding of behavioral-epidemic interactions and offer cautionary insight into the design of intervention strategies.

\section{Model and Methods}

\subsection{Epidemic Dynamics with Behavior-Based Transmission}

We use a discrete-time SIR (Susceptible–Infected–Recovered) model implemented on a static undirected network. Each node represents an individual and can be in one of three epidemiological states: susceptible (S), infected (I), or recovered (R). At each time step, infected individuals attempt to transmit the disease to their susceptible neighbors, and then recover with fixed probability.

To model behavioral heterogeneity, we divide the population into two equally sized groups:
\begin{itemize}
    \item \textbf{Compliant} nodes: those who adhere to anti-epidemic measures (e.g., mask-wearing, isolation, vaccination);
    \item \textbf{Non-compliant} nodes: those who ignore such measures.
\end{itemize}

We introduce two parameters that control the effect of behavior on transmission probability:
\begin{itemize}
    \item \( \delta \in [0, 1] \): scales the infectiousness of compliant infected nodes (eg. impact of mask wearing);
    \item \( \eta \in [0, 1] \): scales the susceptibility of compliant susceptible nodes (eg. impact of vaccination).
\end{itemize}

Given a baseline transmission probability \( p \), the probability of disease transmission along an edge depends on the behavioral types of both individuals involved:

\begin{itemize}
    \item \textbf{Non-compliant (I) → Non-compliant (S)}: transmission with probability \( p \);
    \item \textbf{Compliant (I) → Non-compliant (S)}: probability \( \delta p \);
    \item \textbf{Non-compliant (I) → Compliant (S)}: probability \( \eta p \);
    \item \textbf{Compliant (I) → Compliant (S)}: probability \( \delta\eta p \).
\end{itemize}

All infected nodes recover independently with fixed probability \( r \). This recovery probability is the same across both groups.

Each simulation begins with a fixed number of initially infected nodes, randomly selected from both groups. The epidemic evolves until no infected nodes remain. We record the final fraction of nodes in the \textit{Recovered} state within each behavioral group.

\subsection{Network Models}

To study the interplay between network structure and behavioral homophily in epidemic spreading, we generate synthetic social networks using two complementary approaches: the Stochastic Block Model (SBM) and a network model based on Triadic Closure (TC). Both approaches allow us to manipulate the level of group separation (homophily) while preserving the average degree of the network, ensuring that any differences in epidemic outcomes are attributable to structural and behavioral factors, not changes in connectivity.

\paragraph{Stochastic Block Model (SBM).} In the SBM, nodes are divided into two groups---compliant and non-compliant---each containing \( N/2 \) nodes. Edges are placed probabilistically between node pairs according to their group membership. We define the connection probability matrix:
\begin{equation}
P = \rho \cdot
\begin{pmatrix}
\displaystyle \frac{N(N-1) - 2a N_1 N_2}{N_1(N_1 - 1) + N_2(N_2 - 1)} & a \\
 a & \displaystyle \frac{N(N-1) - 2a N_1 N_2}{N_1(N_1 - 1) + N_2(N_2 - 1)}
\end{pmatrix},
\end{equation}
where \( \rho \) is the desired global network density, \( a \in [0,1] \) controls the level of between-group connectivity (asymmetry), and \( N_1 = N_2 = N/2 \) are the sizes of the two groups. This formulation ensures that the expected total number of edges remains constant regardless of \( a \), allowing us to isolate the effect of homophily.

When \( a = 1.0 \), connections are distributed randomly across the whole population (fully mixed network). As \( a \) decreases, within-group edges become more probable, representing increasing homophily and social separation between compliant and non-compliant individuals. At \( a = 0 \), the network becomes fully bipartitioned with no inter-group edges.


\paragraph{Triadic Closure (TC).} The TC model \cite{raducha2017axelrod} is designed to capture an essential feature of real social networks: local clustering. We begin by generating an Erd\H{o}s--R\'enyi (ER) random graph with \( N \) nodes and average degree \( \langle k \rangle \), and then iteratively rewire edges to increase clustering without changing the number of edges.

The TC construction proceeds in two phases:
\begin{enumerate}
    \item \textbf{Termalization.} Randomly select an edge and replace it with an edge that closes a triangle by connecting a random endpoint to a second-order neighbor. Repeat this process until the global clustering coefficient reaches a plateau.
    \item \textbf{Behavioral modularity.} Assign nodes to two groups of equal size (compliant and non-compliant), then further rewire edges to increase the proportion of intra-group connections. This is done by selecting edges and reconnecting them to second-order neighbors of random endpoint but only if it belongs to the same group, gradually increasing modularity while maintaining the total number of edges. For selected asymmetry, the process ends when 
\begin{equation}
1-2 \times \mbox{graph modularity} < \mbox{asymmetry}
\end{equation}
\end{enumerate}

This two-step rewiring algorithm results in networks with nontrivial clustering and tunable behavioral homophily. Importantly, the procedure preserves the average degree and network size, enabling meaningful comparisons with SBM.

In both models, the parameter \( a \) serves as a homophily control: \( a = 1.0 \) denotes full mixing, while \( a = 0.1 \) represents strong separation between groups. Although \( a \) does not yield the same graph modularity in TC and SBM, it monotonically tunes the group separation in both models
Comparing epidemic outcomes across SBM and TC networks for the same \( a \) and transmission parameters allows us to isolate the role of local clustering in shaping infection dynamics.

\subsection{Simulation Procedure}

We simulate epidemic outbreaks on undirected networks of \( N = 5000 \) nodes, evenly split into two behavioral groups: \textbf{compliant} and \textbf{non-compliant}, each with 2500 nodes. The average degree is fixed at \( \langle k \rangle = 4 \), ensuring consistent network density across all scenarios. We focus on a baseline transmission probability \( p = 0.05 \) and a recovery probability \( r = 0.2 \) for all infected individuals.

Each simulation begins by randomly infecting 50 compliant and 50 non-compliant nodes. All remaining nodes are initially in the susceptible state \( S \). The simulation evolves in discrete synchronous time steps. The disease spreads solely through direct contact with infected neighbors; there is no environmental transmission or memory of prior states. Each realization proceeds until the absorbing state is reached (i.e., the set of infected nodes is empty).

To account for randomness in both the network generation and the initial conditions, we conduct \textbf{1000 independent simulations} for each parameter configuration. In every run a new network is generated according to the selected model (SBM or TC) and asymmetry level \(a\).

We explore the full grid of parameter combinations: \( \delta, \eta \in \{1.0, 0.50, 0.20, 0.10, 0.05, 0.02, 0.01\} \). The results are aggregated by averaging across realizations and by analyzing the full distribution of outcomes within each condition, enabling robust conclusions about trends and variability in epidemic size and direction of transmission.

\section{Results}

We begin our analysis by investigating how the final epidemic size varies between the compliant and non-compliant groups, as a function of behavioral homophily. Specifically, we focus on the fraction of nodes in the \textit{Recovered} state at the end of the simulation, separately for each group. This provides a measure of the total epidemic burden experienced by compliant and non-compliant individuals under different network and behavioral configurations.
\begin{figure}[htbp]
    \centering
    \includegraphics[width=0.4\textwidth]{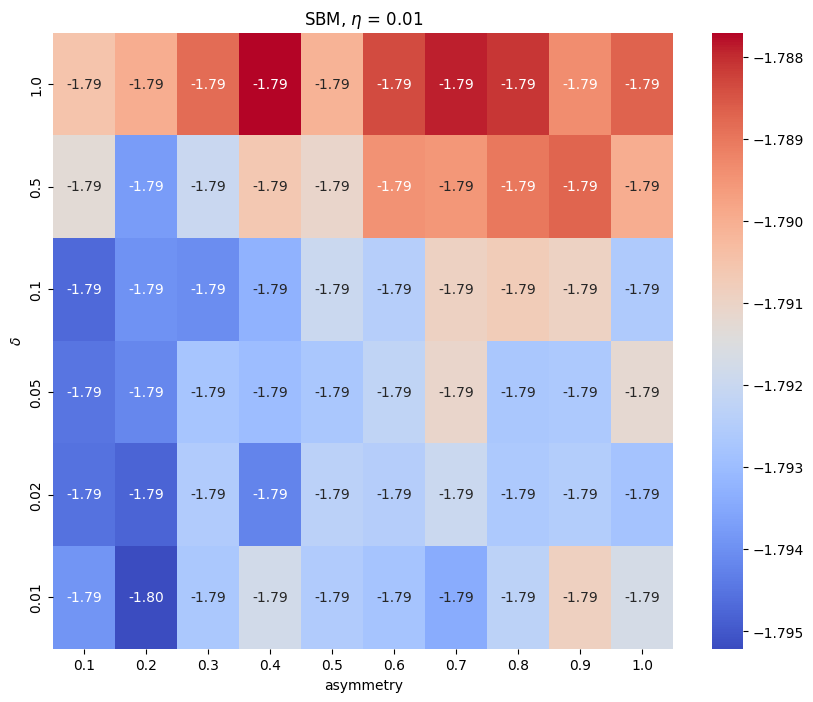}
    \includegraphics[width=0.4\textwidth]{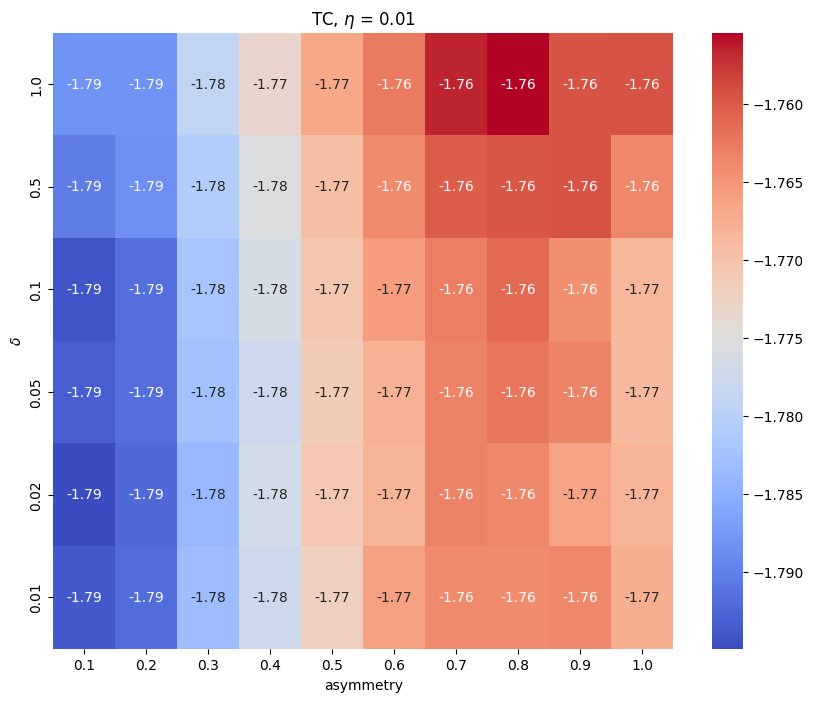}
    \includegraphics[width=0.4\textwidth]{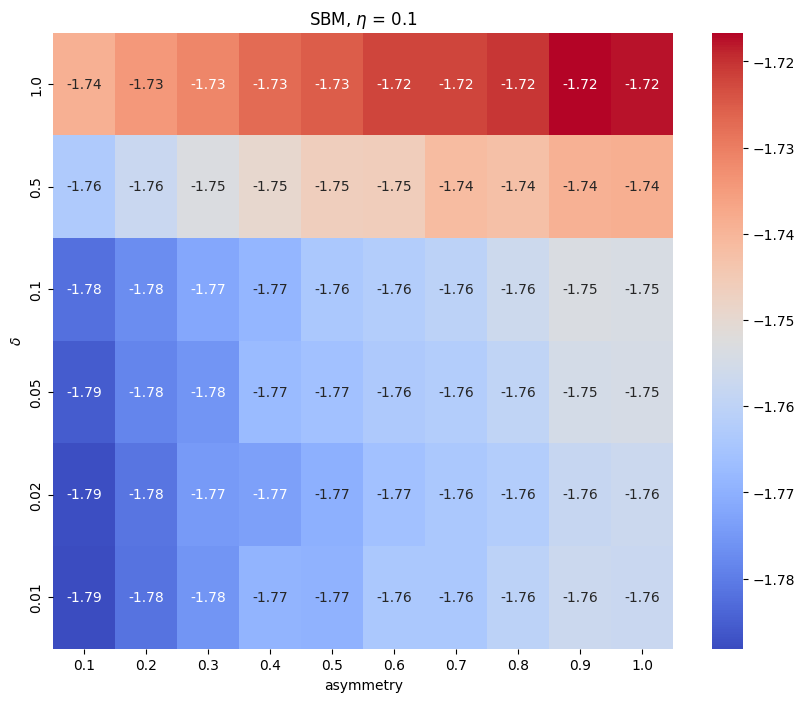}
    \includegraphics[width=0.4\textwidth]{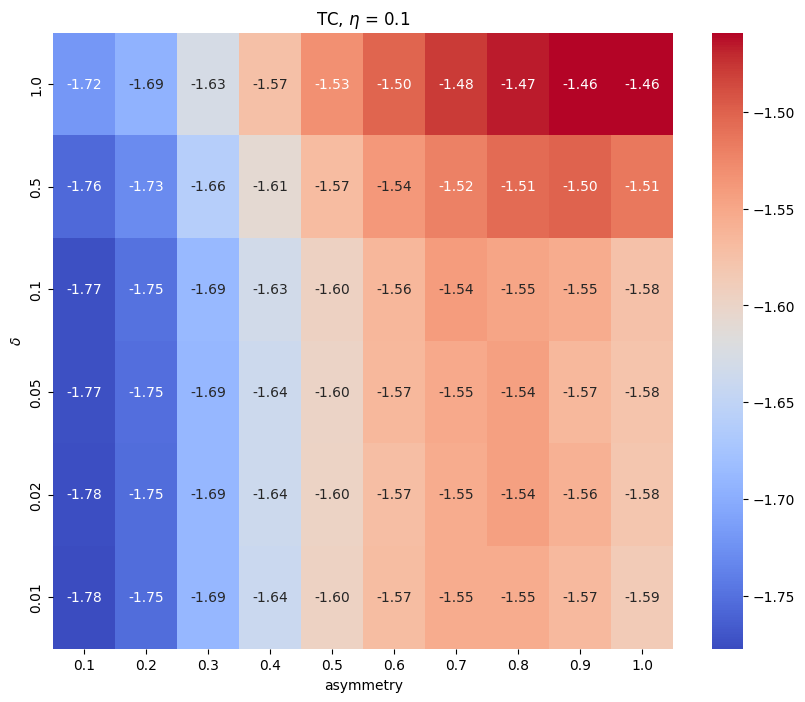}
    \includegraphics[width=0.4\textwidth]{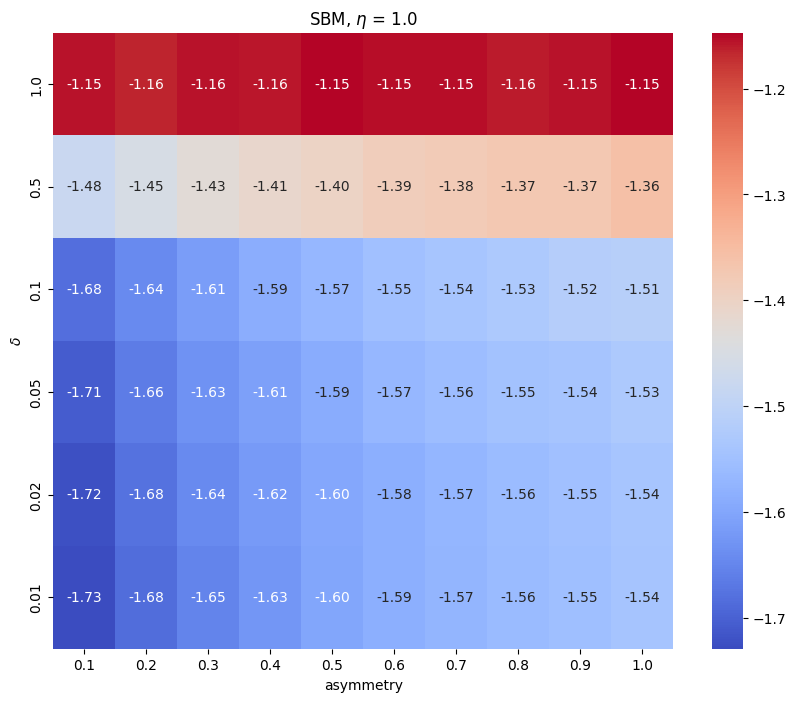}
    \includegraphics[width=0.4\textwidth]{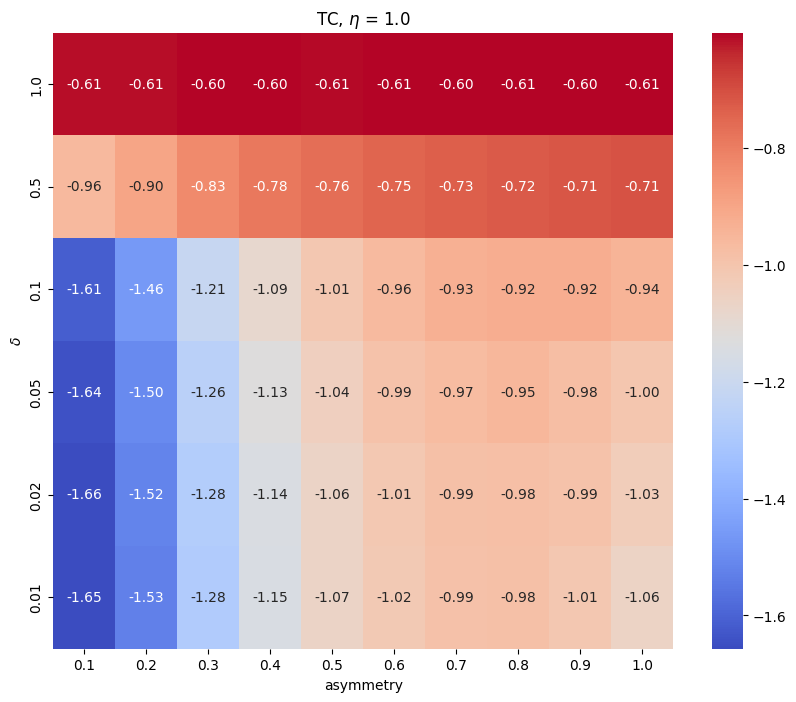}
    \caption{Logarithm (base 10) of the expected fraction of recovered nodes in the compliant group, plotted as a function of group asymmetry, for various values of the susceptibility-scaling parameter \(\eta\). Top three plots correspond to TC networks; bottom three to SBM. All simulations use \(\delta = 0.01\). The non-monotonic increase in epidemic size with asymmetry—most pronounced for \(\eta = 1\)—is visible in all TC cases, but is absent in SBM.}
    \label{fig:log10_compliant_recovered}
\end{figure}
Figure~\ref{fig:log10_compliant_recovered} presents the logarithm (base 10) of the expected fraction of recovered nodes in the compliant group, shown for various combinations of \(\delta\), \(\eta\), and asymmetry levels, separately for TC and SBM networks.

A key observation is that for small values of \( \delta \) (e.g., \( \delta = 0.01, 0.02 \)), the fraction of recovered nodes in the compliant group \textit{initially increases} with group separation, reaching a maximum around \( a \approx 0.7 \), before decreasing again. 
Analogous results for the non-compliant group are consistently monotonic across all parameters and network types; for brevity, we do not include them here.
\begin{figure}[ht]
    \centering
    \includegraphics[width=0.9\textwidth]{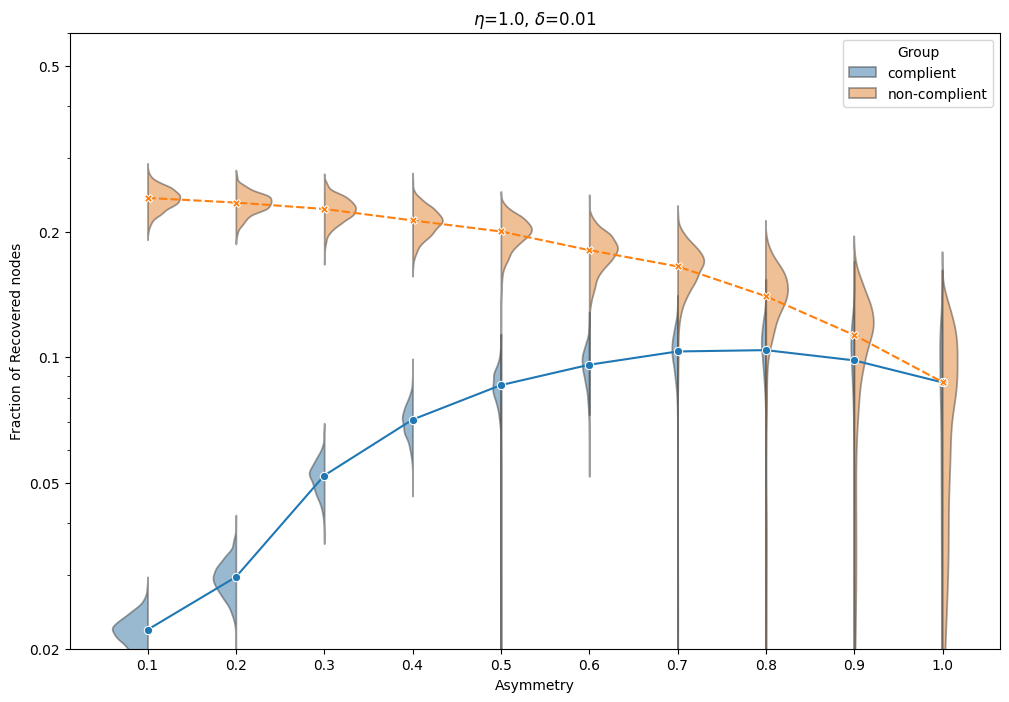}
    \caption{Distribution and mean fraction of recovered nodes in compliant and non-compliant groups as a function of asymmetry, for the Triadic Closure (TC) network model. Parameters: \( \eta = 1.0 \), \( \delta = 0.01 \). Note the non-monotonic behavior in the compliant group, with a peak around asymmetry \( a \approx 0.7 \).}
    \label{fig:tc_rfrac}
\end{figure}
Figure~\ref{fig:tc_rfrac} shows the distribution and mean fraction of recovered nodes in each group for the TC model, across a range of asymmetry levels \( a \), and for transmission asymmetry parameters \( \eta = 1.0 \), \( \delta = 0.01 \) fixed.
This non-monotonic behavior is unexpected: stronger separation from the high-risk non-compliant group intuitively suggests better protection. However, the increased clustering among non-compliant individuals accelerates within-group transmission and leads to more secondary infections that eventually reach compliant individuals via residual inter-group links.

This paradoxical effect is most pronounced when \( \eta = 1.0 \), but it persists for smaller values of \( \eta \) as well. Importantly, the effect emerges consistently when \( \delta \ll 1 \), regardless of \( \eta \), suggesting that it is the asymmetry in \textit{infectiousness}—not susceptibility—that drives the non-monotonic outcomes.

By contrast, in the non-compliant group, the fraction of recovered nodes increases monotonically with homophily: as separation strengthens, their denser internal connectivity facilitates more efficient spread.

For larger values of \( \delta \) (e.g., \( \delta = 0.5, 1.0 \)), the asymmetry in infectiousness disappears, and both groups exhibit nearly identical epidemic sizes that vary little with \( a \).

In this case of SBM, the non-monotonic effect observed in TC is absent. In the compliant group, the epidemic size either decreases monotonically or remains flat as \( a \) decreases. This confirms that the paradoxical effect seen in TC arises from the interaction between clustering and behavioral heterogeneity.

These findings highlight that network structure plays a crucial role in shaping the impact of behavioral homophily on epidemic dynamics. In particular, clustering-induced amplification of internal transmission within the non-compliant group can counteract the protective effect of separation, resulting in increased risk for the compliant group.

\section{Discussion and Conclusions}

\subsection{Implications of Homophily for Epidemic Control}

Our findings reveal a nuanced and at times counterintuitive role of behavioral homophily in shaping epidemic outcomes. While separating compliant individuals from non-compliant ones may intuitively appear protective, our results show that this strategy can fail—or even backfire—under specific conditions.

In clustered networks such as those generated by the Triadic Closure (TC) model, increasing homophily amplifies internal connectivity within each group. When the asymmetry in transmission is large (i.e., \( \delta \ll 1 \)), non-compliant individuals become the dominant vectors of infection. As internal transmission accelerates within the non-compliant group, the residual inter-group links can serve as bridges, enabling infection to spill over into the compliant group. Paradoxically, this means that partial separation (e.g., \( a = 0.7 \)) can result in a higher epidemic burden for compliant individuals than no separation (\( a = 1 \)).

Interestingly, our results show that this paradoxical effect occurs across all examined values of \( \eta \), although it is most pronounced when \( \eta = 1 \). This indicates that the effect is not driven by the susceptibility of compliant individuals, but by the infectiousness of compliant infecteds—that is, by \( \delta \). Since \( \delta \) represents the effectiveness of interventions that reduce outward transmission (e.g., mask-wearing), our findings suggest that clustering and partial separation can undermine such measures even when they are highly effective at source control. In contrast, interventions that affect \( \eta \) (e.g., vaccines reducing susceptibility) do not produce this paradoxical behavior.

This insight aligns with recent studies such as \cite{klimek2024maskshomophily}, which show that the effectiveness of masks may paradoxically increase when fewer people wear them, due to interaction with network structure and behavior clustering. Similarly, \cite{are2024vaccinehomophily} found that homophily in vaccination status raises the infection risk among unvaccinated individuals, even when overall coverage is high.

Crucially, the paradoxical pattern emerges only under three conditions: strong behavioral asymmetry (\( \delta \ll 1 \)), high clustering (as in TC networks), and partial—not complete—group separation. It does not occur in random-like networks generated by the Stochastic Block Model (SBM), nor when transmission asymmetry is weak (\( \delta \approx 1 \)).

These results suggest that segregation-based mitigation strategies, whether spontaneous or policy-driven, should be evaluated not only based on their local behavioral logic, but also in terms of their systemic effects in clustered social structures. Failure to account for network topology may lead to unintended consequences.

\subsection{Limitations and Future Work}

While our model provides valuable qualitative insights, several limitations must be acknowledged.

First, we rely on synthetic networks generated by two stylized models: SBM and TC. Although these capture key features like modularity and clustering, real-world contact networks also involve degree heterogeneity, overlapping communities, and temporal evolution. Recent advances in higher-order network modeling \cite{burgio2024triadicapprox,malizia2025gbcm,contreras2024patterns} could help extend our framework to more realistic settings.

Second, behavior is modeled in a binary and static manner: individuals are either compliant or non-compliant, with fixed \( \delta \) and \( \eta \). In practice, adherence may be probabilistic, context-dependent, or change over time due to feedback from epidemic progression \cite{alutto2025nbf}. Incorporating dynamic behavioral models and feedback mechanisms would be a promising avenue for future work.

Third, our focus is on the final epidemic size and directional transmission counts, rather than time-resolved dynamics. While this is sufficient to uncover structural effects, it limits interpretation for real-time intervention planning.

Fourth, we assume infections occur only via direct links, with no memory or long-range spreading. While consistent with many SIR-type models, this abstraction may overlook mechanisms such as superspreading events or mobility-driven contacts.

Finally, although our results are not predictive, they offer conceptual clarity about when and why behavioral clustering may have unintended epidemiological consequences. Future research may explore multilayer networks \cite{bellingeri2024perspective}, vaccination feedback, or policy scenarios involving hybrid NPI strategies.

Our work underscores that effective epidemic mitigation depends not only on individual behavior but also on how behavior is embedded in network structure—a point increasingly recognized across epidemiology, sociology, and network science \cite{bizzarri2024homophilyeffects,cirigliano2024stc}.

\section*{Acknowledgements}
The authors gratefully acknowledge discussions with colleagues at the Faculty of Physics and the Faculty of Artes Liberales, University of Warsaw.

\section*{Funding}
This research was founded by the Polish Ministry of Science and Higher Education's "Initiative for Excellence — Research University" programme, granted to the University of Warsaw.

\section*{Conflicts of Interest}
The authors declare no conflict of interest.

\bibliographystyle{plain}
\bibliography{ref2025}

\end{document}